\documentclass[aps,prb,twocolumn,showpacs,prl,superscriptaddress,amsfonts,amsmath,amssymb]{revtex4-1}

\usepackage{graphicx}

\begin{document}

\title{Spin-selective localization of correlated lattice fermions}

\author{J. Skolimowski}
\email{jan.skolimowski@fuw.edu.pl}
\affiliation{Institute of Theoretical Physics,
Faculty of Physics,  University of Warsaw, ul.~Pasteura 5, PL-02-093 Warszawa, Poland}

\author{D. Vollhardt}
\affiliation{Theoretical Physics III, Center for Electronic Correlations and Magnetism,
Institute of Physics, University of Augsburg, D-86135, Augsburg, Germany}

\author{K. Byczuk}
%\email{byczuk@fuw.edu.pl}
\affiliation{Institute of Theoretical Physics,
Faculty of Physics,  University of Warsaw, ul.~Pasteura 5, PL-02-093 Warszawa, Poland}

\date{\today}

\begin{abstract}

The interplay between local, repulsive interactions and disorder acting only on one spin orientation of lattice fermions (''spin-dependent disorder")  is investigated. The nonmagnetic disorder vs. interaction phase diagram is computed using Dynamical Mean-Field Theory in combination with the geometric average over disorder. The latter determines the typical local density of states and is therefore sensitive to Anderson localization. The effect of spin-dependent disorder is found to be very different from that of conventional disorder. In particular, it destabilizes the metallic solution and leads to a novel spin-selective, localized phase at weak interactions and strong disorder.

\end{abstract}


\maketitle
\section{Introduction}

Experiments with optically controlled cold atoms are able to turn new concepts and  theoretical models into reality.\cite{bloch08} For example, the Bose-Hubbard Hamiltonian\cite{jaksch98} was realized and the superfluid-insulator transition was observed,\cite{greiner02} a fermionic Mott insulator was created,\cite{jordens08}  Anderson localization was detected,\cite{kondov11}  the spin Hall effect was measured,\cite{beeler13} and exotic features such  as fractional statistics were perceived.\cite{keilmann11}
In particular, such experiments make it possible to simulate models and discover phenomena which are absent in solid state physics. One such model, namely  fermions on a lattice  with  spin-dependent hopping,  was already realized.\cite{mandel03} In another experiment a spin-dependent periodic potential was formed.\cite{mckay10} A possible next step is to implement a lattice system with  spin-dependent disorder by combining spin-dependent and local random potentials.\cite{Palencia10}

Spin-dependent disorder may, in principle, be realized experimentally by focusing light beams with different polarization on an optical lattice, after having been scattered from a diffusive plate.\cite{makuch13} This leads to a spin-dependent, speckle-type effective random potential which is characterized by a short correlation length and  pronounced statistical independence\cite{mandel03,mckay10,Palencia10} and which acts differently on particles with different spin orientation in the ground state. Thereby spin-dependent speckle-type disorder modifies the local one-particle energies, hopping amplitudes, and interparticle interaction potentials.

In this paper we  discuss the Anderson and Mott metal-insulator transitions (MIT) within the fermionic Hubbard model in the presence of spin-dependent disorder. Surprisingly, up to now the influence of such a type of disorder on correlated lattice fermions has not yet been investigated in detail. Scalettar and collaborators\cite{scalettar12} analyzed the pairing instability in the attractive case, while the present authors studied the thermodynamical properties in the repulsive case.\cite{makuch13} By employing the dynamical mean-field theory (DMFT) with arithmetic average over the disorder it was shown\cite{makuch13} that  spin-dependent disorder with a symmetric probability distribution function (PDF), $\mathcal{P}_{\sigma}(\epsilon) = \mathcal{P}_{\sigma}(-\epsilon)$, yields a finite magnetization in the repulsive Hubbard systems on a bipartite lattice away from half-filling. At half-filling this novel type of disorder breaks the particle-hole and spin symmetries, but no magnetization appears. The absence of magnetization is explained by the fact that the symmetry of the spin resolved local density of states (LDOS) $\rho_{\sigma}(\omega) = \rho_{\sigma}(-\omega)$ is preserved in the interacting system, so that the number of states below the Fermi energy $\omega=0$ is unchanged.

 There remains the  questions how spin-dependent disorder will affect the Mott-Hubbard transition and Anderson localization, respectively. In particular, in the case of non-interacting fermions, particles with different spin are independent and Anderson localization can therefore occur only in that spin subsystem on which the disorder acts. In the following we will shown that when the interaction is turned on, metallic and  Anderson  insulating states for different spin subsystems coexist in a large part of the phase diagram.

For this purpose we employ the DMFT together with the \emph{geometric} average over the disorder, since the latter is sensitive to Anderson localization even within a one-particle description.\cite{Dobrosavljevic97,Dobrosavljevic03,Schubert03} Employing the geometric rather than the arithmetic average corresponds to determining the typical LDOS\cite{Anderson58}
within the DMFT. This approach was recently used to calculate the paramagnetic\cite{Byczuk05} and antiferromagnetic\cite{Byczuk09} phase diagrams of the disordered Hubbard model as well as the phase diagram of the disordered Falicov-Kimball model.\cite{byczuk05a} Thereby it was possible to determine the MIT due to disorder (Anderson localization) and interactions (Mott-Hubbard transition), respectively,
as well as the rich transition scenario caused by the simultaneous presence of interactions and disorder,
within a unified framework. A recent experiment on correlated, disordered fermionic cold atoms\cite{Kondov15} confirmed that the Anderson localization line indeed increases linearly with the interaction strength as predicted in Ref.~\onlinecite{Byczuk05}.

In the following we will refer to disorder acting equally on both spin subsystems as ''conventional disorder`` to distinguish it from the spin-dependent disorder discussed here, where different random one-particle potentials act on the individual spin subsystems. In this study we do not consider long-range antiferromagnetic order since current experiments are still at temperatures above the Neel temperature.\cite{Hart15}

\section{Model and method}

The system under consideration  is modeled by the Anderson-Hubbard Hamiltonian at half filling with spin-dependent diagonal disorder
\begin{eqnarray}
\label{AH}
H =  \sum_{ ij \sigma} t_{ij} a^\dagger_{i\sigma}a_{j\sigma}+\sum_{i\sigma}\epsilon_{i\sigma} n_{i\sigma}+ \nonumber \\
U\sum_i \left( n_{i\uparrow}-\frac{1}{2}\right) \left(n_{i\downarrow}-\frac{1}{2}\right) ,
\end{eqnarray}
where $a_{i\sigma}$ ($a^\dagger _{i\sigma}$) is the annihilation (creation) operator of a fermion at site $i$, $\sigma =\pm 1/2=\uparrow,\;\downarrow$ is the $z$ component of the spin, $n_{i\sigma}=a^\dagger_{i\sigma}a_{i\sigma}$ is the particle number operator, $t_{ij}$ is the hopping amplitude between site $i$ and $j$, and $U$ is the on-site repulsion.

The local spin-dependent  potentials $\epsilon_{i\sigma}$ represent  uncorrelated random variables drawn from a PDF  $\mathcal{P}_{\sigma}(x)$. In this study the spin-dependent disorder is modeled by a rectangular  (box)  PDF  given by
\begin{equation}
\mathcal{P}_{\sigma}(x)=\frac{1+2\sigma}{2 \Delta}\Theta\left(\frac{\Delta}{2}-|x|\right),
\end{equation}
where  $\Theta (y)$ is  the Heaviside step function and $\Delta$ is the strength of the   disorder. It means that particles with spin $\sigma =\uparrow$ move on a lattice with a random potential whereas  particle with spin $\sigma =\downarrow$ propagate on a uniform lattice.
In the absence of disorder the system has SU(2) spin symmetry and, at half-filling,  is also particle-hole symmetric. The spin-dependent disorder breaks both symmetries since the on-site energy $\epsilon_{i\sigma}$  is equivalently represented  by a random local chemical potential $\mu_i=-(\epsilon_{i\uparrow} + \epsilon_{i \downarrow})/2$ and  a random local Zeeman magnetic field $h_i=(\epsilon_{i\uparrow}-\epsilon_{i\downarrow})/2$.\cite{makuch13}

The Anderson-Hubbard  Hamiltonian (\ref{AH}) is solved within the DMFT.\cite{georges96} Here we map the Hamiltonian (\ref{AH}) onto an ensemble of single-impurity Anderson models
\begin{eqnarray}
H_{\rm SIAM} = \sum_{\sigma} \epsilon_{\sigma} n_{\sigma} + U n_{\uparrow} n_{\downarrow} + \nonumber \\
\sum_{{\bf k}\sigma} (V_{{\bf k}\sigma} a^{\dagger}_{\sigma} c_{{\bf k} \sigma} + V_{{\bf k}\sigma }^*  c_{{\bf k} \sigma}^{\dagger} a_{\sigma}) + \sum_{{\bf k}\sigma} E_{{\bf k}\sigma} c_{{\bf k} \sigma}^{\dagger} c_{{\bf k} \sigma}
\label{AIM}
\end{eqnarray}
with random spin-dependent atomic energies $\epsilon_{\sigma}$ drawn from the same PDF $\mathcal{P}_{\sigma}(\epsilon_{\sigma})$ as in (\ref{AH}).
The hybridization matrix elements $V_{{\bf k}\sigma}$ and the dispersion relation $E_{{\bf k}\sigma} $ of the bath fermions $c_{{\bf k} \sigma}$ define the hybridization function $\eta_{\sigma} (\omega) = \sum_{\bf k} |V_{{\bf k}\sigma} |^2/ (\omega - E_{{\bf k}\sigma})$ and are determined self-consistently in the following way:
For each $\epsilon_{\sigma}$ we calculate the impurity Green function $G_{\sigma}(\omega, \epsilon_{\sigma})$, corresponding to (\ref{AIM}), and the LDOS $\rho_{\sigma}(\omega,\epsilon_{\sigma})=-\rm{Im} G_{\sigma}(\omega, \epsilon_{\sigma}) /\pi $. From this we obtain the geometrically averaged LDOS $\rho_{\sigma}(\omega) = \rm{exp} [ \langle \rm{ln} \rho_{\sigma} ( \omega,\epsilon_{\sigma} ) \rangle ]$, where $ \langle Q \rangle = \int d \epsilon \mathcal{P} _{\sigma}(\epsilon) Q(\epsilon)$ denotes the arithmetic average of $ Q(\epsilon) $.
%In  parallel, the arithmetically averaged LDOS $\rho_{\sigma}^{\rm arith} (\omega) = \langle\rho_{\sigma}(\omega,\epsilon_{\sigma}) \rangle $ is also determined.
The local Green function, averaged over the ensemble, is determined by  the Hilbert transform $G_{\sigma} (\omega) = \int d \omega ' \rho_{\sigma} (\omega ')/(\omega - \omega ')$.
%, where $\alpha=\rm{geom}$ (arith) marks the type of averaging used in computing LDOS.
The local self-energy $\Sigma _{\sigma} (\omega) $ is then obtained form the $\bf k$-integrated Dyson equation $ \Sigma _{\sigma}(\omega) = \omega - \eta_{\sigma}(\omega) - 1 / G_{\sigma}(\omega)$, where $\eta_{\sigma} (\omega) $  describes  the mean coupling of the single lattice site to the rest of the system in the DMFT. The DMFT equations are closed by the Hilbert transform $G_{\sigma}(\omega) = \int d \xi N_0(\xi) / [ \omega - \xi - \Sigma _{\sigma} (\omega ) ] $, where $N_0(\xi)$ is the noninteracting density of states (DOS), which relates the local Green function for a given lattice to the self-energy from the impurity model. These equations must be solved iteratively to reach self-consistency. For comparison we also calculate the arithmetically averaged LDOS $\rho_{\sigma}^{\rm arith}(\omega) = \langle \rho_{\sigma} ( \omega,\epsilon_{\sigma} ) \rangle $.

We choose a model DOS, $N_0(\xi) = 2 \sqrt{D^2-\xi^2}/\pi D^2$, where $W=2D$ is the bandwidth, and $W=1$ sets the energy unit. For this DOS the local Green function and the hybridization function are simply related by $G_{\sigma}(\omega)= D^2 \eta_{\sigma}(\omega)/4$.\cite{georges96} Here, the system is assumed to be paramagnetic down to the zero temperature. The phase diagram and dynamical quantities for the model (\ref{AH}) are computed  by solving the DMFT equations at zero temperature  with the numerical renormalization group method (NRG)\cite{wilson,bulla2008} in each iteration step. For this purpose we implemented the open source NRG Ljubljana code.\cite{nrgljubljana}

\section{Results}

\begin{figure}[tbp]
	\begin{center}
		\includegraphics[clip,width=0.5\textwidth]{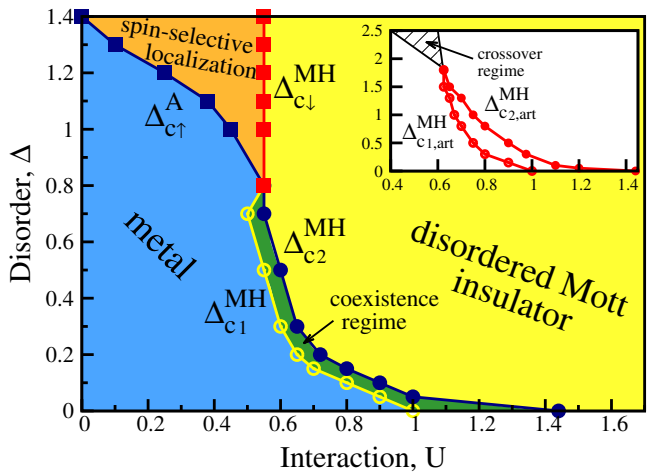}
		\caption{Nonmagnetic  ground state phase diagram of the Anderson-Hubbard model at half filling  with spin-dependent disorder determined by DMFT with the typical local density of states (LDOS). The inset shows the phase diagram obtained from the arithmetically  averaged LDOS. }
		\label{phase_diagram}
	\end{center}
\end{figure}

The main result of our investigation is the ground state phase diagram of the Anderson-Hubbard model with  spin-dependent disorder shown in Fig.~\ref{phase_diagram}. It differs significantly from the corresponding phase diagram for conventional disorder.\cite{Byczuk05} Three different phase transitions take place: (i) a Mott-Hubbard type MIT for weak disorder $\Delta$, where the correlation gap  at the Fermi level opens in the LDOS for spin up and down, (ii) an  Anderson MIT for weak interaction $U$, where the LDOS for spin up vanishes, and (iii) a Falicov-Kimball type MIT, where the DOS for spin down acquires a correlation gap. The correlated, disordered metallic phase borders on the Mott insulator and the Anderson localized phase, where the spin-up particles are localized while those with spin down are still itinerant. In the following we will refer to the latter phenomenon as ''spin-selective localization``. The novel spin-selective localized phase originates from the spin dependence of the disorder. In the following we describe the properties of these three phases and characterize the transitions between them in detail.

\subsection{Metallic phase}

\begin{figure}[tbp]
	\begin{center}
 		\includegraphics[clip,width=0.5\textwidth]{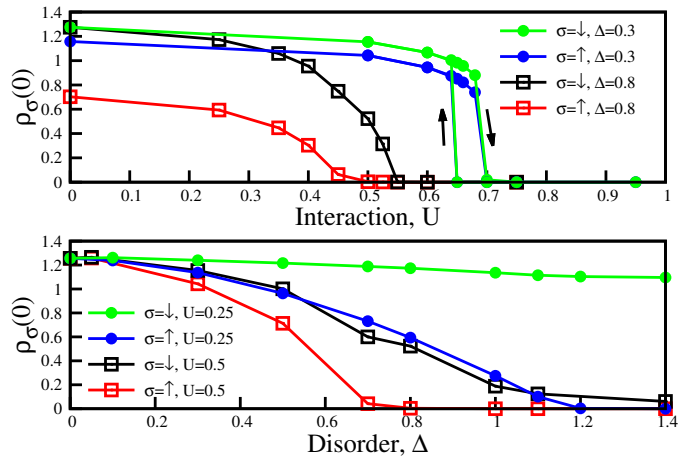}
		\caption{Local density of states at the Fermi level as a function of interaction strength $U$  (upper panel) and disorder $\Delta$ (lower panel).}
		\label{dos_zero}
	\end{center}
\end{figure}

The correlated metallic phase is defined by a non-zero value of the geometrically averaged LDOS at the Fermi level, $\rho_{\sigma} (0)\neq 0$, for both spin channels. In the metallic phase without disorder ($\Delta=0$) $\rho_{\sigma} (0)$ corresponds to the non-interacting DOS $N_0(0)$ for both spins and arbitrary $U $. In this case the Luttinger theorem is obeyed and Landau quasiparticles at the Fermi level are well defined.\cite{georges96}  In the presence of  spin-dependent disorder, $\rho_{\sigma}(0)$ is reduced for particles in both spin channels, but differently depending in the spin direction. Only for $U=0$ are the spin down particles not affected by the disorder at all. The reduction of  $\rho_{\sigma}(0)$  due to an increase of $\Delta$ or $U$ is shown in Fig.~\ref{dos_zero}. In contrast to the case of conventional disorder\cite{Byczuk05}  where a sufficiently strong interaction was found to protect the quasiparticles from decaying due to impurity scattering, $\rho_{\sigma}(0)$ now always decreases with $U$ for any finite $\Delta$. Spin-dependent disorder  violates the Luttinger pinning condition at any $U$, and the local interaction cannot restore the Landau quasiparticle picture. As we will discuss later, this originates from the SU(2) spin symmetry breaking by this type of disorder.

\subsection{Mott-Hubbard MIT and the coexistence regime}

\begin{figure}[tbp]
	\begin{center}
		\includegraphics[clip,width=0.5\textwidth]{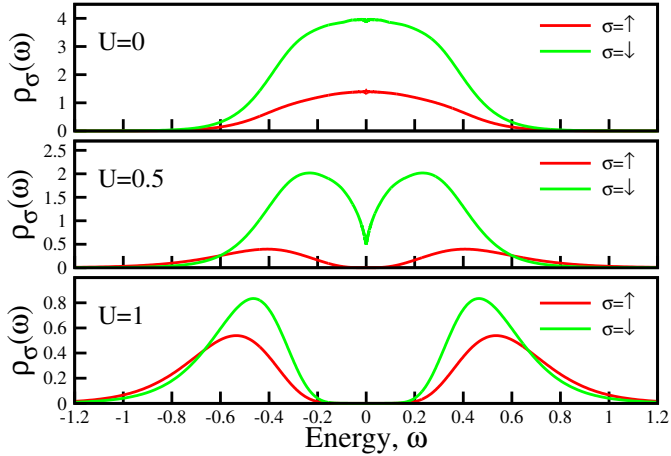}
		\caption{Local density of states for different values of the interaction U at disorder strength $\Delta=0.5$. Upper panel: disordered metallic phase away from the transition line $\Delta^{\rm{MH}}_{\rm{c}2}$; middle panel: disordered metallic phase close to the transition line $\Delta^{\rm{MH}}_{\rm{c}2}$; lower panel: disordered Mott insulator at the critical interaction strength $U=0.6$.}
		\label{hubbard}
	\end{center}
\end{figure}

 In the Mott insulating phase the LDOS vanishes at the Fermi level, $\rho_{\sigma} (0)= 0$. This is seen in the upper panel in Fig.~\ref{dos_zero}, which leads to Hubbard subbands for stronger interactions U, cf. Fig.~\ref{hubbard}.  The MIT at weak disorder, $\Delta\lesssim 0.8$, is accompanied with hysteresis (upper panel in Fig.~\ref{dos_zero}), and the transition lines $\Delta^{MH}_{c1}(U)$ and $\Delta^{MH}_{c2}(U)$ are tilted to the left (Fig.~\ref{phase_diagram}). These transition  lines terminate at a single critical point close to $U\approx 0.6$ and $\Delta\approx 0.8$.
In contrast to conventional disorder,\cite{Byczuk05} where the critical point is located at $ \Delta\approx 1.8$, the coexistence regime is found to be significantly smaller, and a  crossover regime  does not  occur at all.
These differences must be  attributed to the reduced spin-symmetry in the present problem. Namely, the quasiparticle central peak between the Hubbard subbands, which originates from spin-flip scattering and thereby leads to the Kondo resonance, is destroyed when a Zeeman magnetic field is applied.\cite{kondo5,kondo4,kondo3b,kondo3a,kondo3,kondo2,kondo1} The induced finite magnetization reduces the number of scattering states at the Fermi level in different spin channels. Although in the present problem the magnetization is zero,\cite{makuch13} the density of states in the two spin bands differ, cf.~Fig.~\ref{hubbard}. Altogether,  spin-dependent disorder reduces the metallicity, forcing the interacting system into an insulating state, in contrast to the spin-independent case.

\begin{figure}[tbp]
	\begin{center}
		\includegraphics[clip,width=0.5\textwidth]{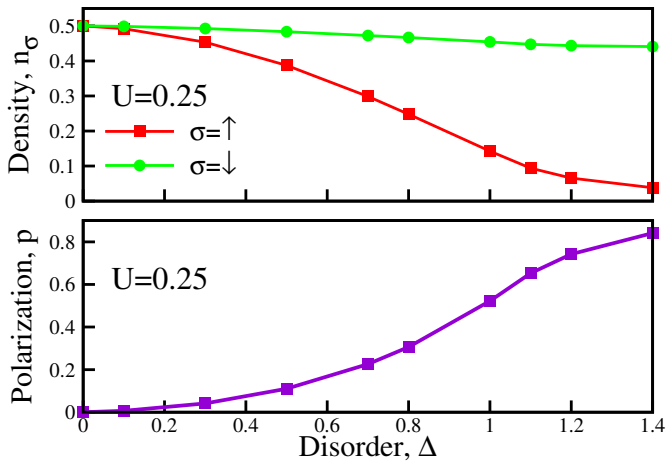}
		\caption{Upper panel: Density of particles with spin up and spin down, respectively, in extended states at $U=0.25$ as a function of disorder strength $\Delta$. Lower panel: Polarization of extended states at $U=0.25$ as a function of disorder.  }
		\label{polarization1}
	\end{center}
\end{figure}

\begin{figure}[tbp]
	\begin{center}
		\includegraphics[clip,width=0.5\textwidth]{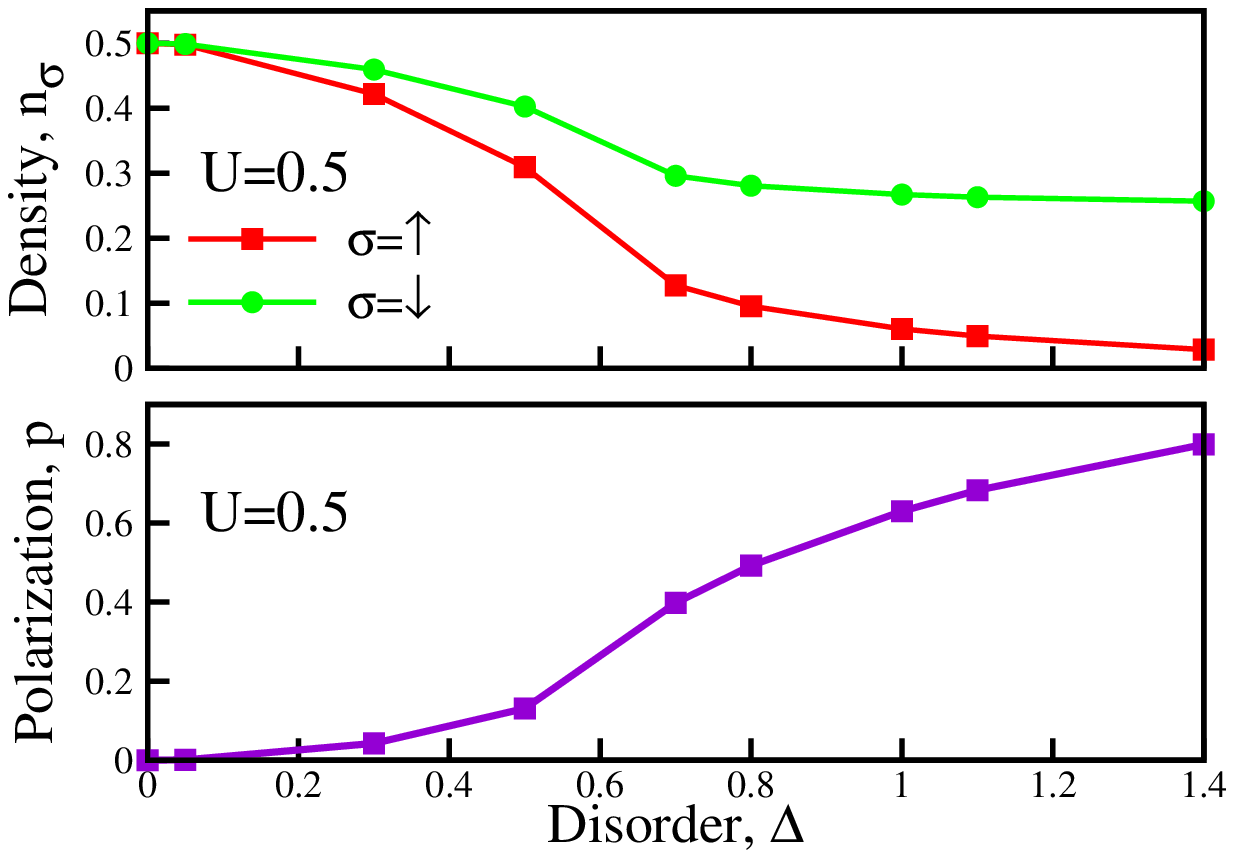}
		\caption{Upper panel: Density of particles with spin up and spin down, respectively, in extended states at $U=0.5$ as a function of disorder strength $\Delta$. Lower panel: Polarization of extended states at $U=0.5$ as a function of disorder.}
		\label{polarization2}
	\end{center}
\end{figure}

\begin{figure}[tbp]
	\begin{center}
		\includegraphics[clip,width=0.5\textwidth]{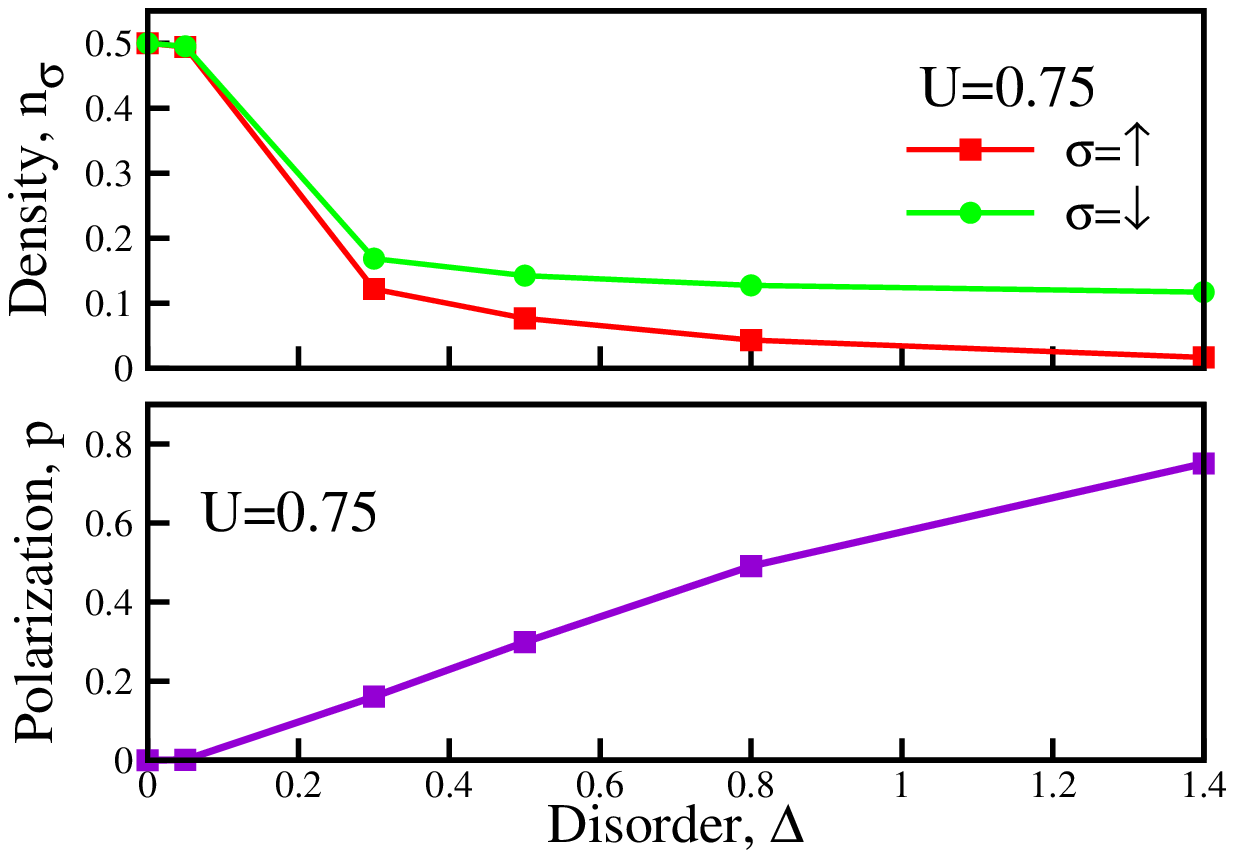}
		\caption{Upper panel: Density of particles with spin up and spin down, respectively, in extended states at $U=0.75$ as a function of disorder strength $\Delta$. Lower panel: Polarization of extended states at $U=0.75$ as a function of disorder.}
		\label{polarization3}
	\end{center}
\end{figure}

\begin{figure}[tbp]
	\begin{center}
		\includegraphics[clip,width=0.5\textwidth]{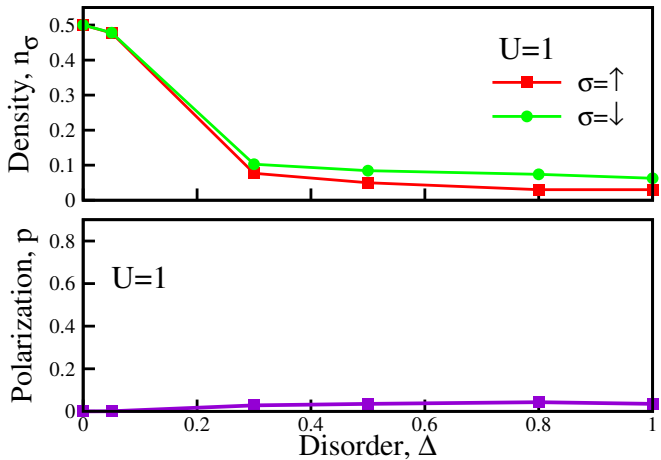}
		\caption{Upper panel: Density of particles with spin up and spin down, respectively, in extended states at $U=1.0$ as a function of disorder strength $\Delta$. Lower panel: Polarization of extended states at $U=1.0$ as a function of disorder.}
		\label{polarization4}
	\end{center}
\end{figure}

 The spin asymmetry of the bands
 %which is induced by spin dependent disorder
  leads to an unequal population of particles with different spin in extended states. This is seen in Figs.~\ref{polarization1}-\ref{polarization4}, where we illustrate the dependence of the density $n_{\sigma}$ of fermions with spins $\sigma$ (upper insets) and of the polarization $p \equiv |(n_{\uparrow}-n_{\downarrow})/(n_{\uparrow}+n_{\downarrow})|$ (lower insets) on the disorder strength $\Delta$ for different interactions $U$.  Without interaction only the spin-up band is affected by the disorder. Due to the interaction between up and down particles the effect of the disorder is transmitted to the spin down band. The difference in the occupation of extended states is reduced by the interaction $U$, which leads to a decrease of the polarization.

\subsection{Spin-selective Anderson localized phase}

In the weakly interacting case, $U\lesssim 0.55$, a spin-selective  Anderson localization is found for $\Delta \geq \Delta^A_{c\uparrow}(U)$, as shown in Fig.~\ref{phase_diagram}. Interestingly, we find that only particles with spin up become localized when $\Delta \geq  \Delta^A_{c\uparrow}(U)$, cf. Fig.~\ref{phase_diagram}, whereas the particles with spin down are still itinerant. The LDOS at the Fermi level is zero in the former case and non-zero in the latter, as  shown in the lower panel of Fig.~\ref{dos_zero} by the blue and green lines, respectively.  On physical grounds this spin-selective localized phase may be effectively interpreted within a Falicov-Kimball model,\cite{Falicov} where spinless lattice fermions interact with immobile particles. In the present case, the particles with spin up are localized due to the disorder $\Delta$ and form an immobile subsystem on which the particles with spin down are scattered due to the interaction $U$. This spin-selective localization implies the absence of  spin-up quasiparticle states around the Fermi level. However, Hubbard subbands are formed in the LDOS for both spins at higher energies, as is shown in the middle panel of Fig.~\ref{hubbard-s}. Again we note that although the disorder acts only on spin-up particles, the interaction transfers the effect of the disorder also to the spin-down particles. This is evident form Figs.~\ref{dos_zero} and \ref{hubbard-s}, where it is seen that some states with spin down become localized. Indeed,  by setting the hybridization function for particles with spin up to zero, one can show that the DMFT action for the Hubbard model reduces to that for the Falicov-Kimball model \cite{Freericks}. This justifies our interpretation of the spin-selective Anderson localized phase in terms of the Falicov-Kimball model.

\subsection{Disordered Mott insulator}

\begin{figure}[tbp]
	\begin{center}
		\includegraphics[clip,width=0.5\textwidth]{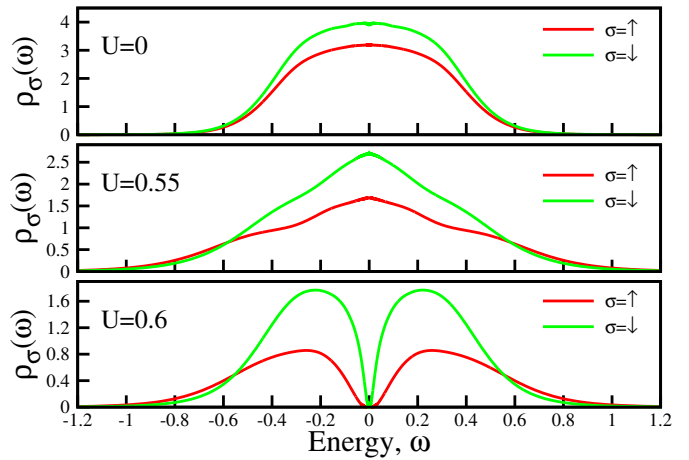}
		\caption{LDOS at disorder strength $\Delta=1$ for different values of the interaction U. Upper panel: disordered metallic phase; middle panel: transition point from the disordered metallic to the spin-selective Anderson localized phase; lower panel: disordered Mott insulator. }
		\label{hubbard-s}
	\end{center}
\end{figure}

 For strong disorder, $\Delta\gtrsim 0.8$, we find a transition from the spin-selective Anderson  localized phase to a disordered Mott insulator upon increasing $U$ ("Falicov-Kimball type MIT"). Namely, there is no hysteresis and the spin-down band is completely split off for $U_c \gtrsim 0.55$. Above this value the LDOS at the Fermi level for spin down particles vanishes, cf. Fig.~\ref{dos_zero}.  The transition line $\Delta^{MH}_{c \downarrow}(U)$ is seen to be vertical for disorder  $\Delta \gtrsim 0.8$. This means that the localized spin-up fermions now play the role of the immobile particles in the Falicov-Kimball model.  By further increasing $U$ for a fixed $\Delta$ a Mott gap proportional to $U$  develops, cf. the lower panel in Fig.~\ref{hubbard-s}. On the insulating side the integral over the LDOS changes with $U$ although $\Delta$ is held constant. This is due to the Hubbard type coupling between spin up and down particles, leading to changes in both hybridization functions $\eta_{\sigma}(\omega)$. Therefore, we conclude that although the MIT above $\Delta\approx 0.8$ is of the Falicov-Kimball type, the phase on the right hand side of the phase boundary in Fig.~\ref{phase_diagram} is still a disordered Mott insulator. In other words, there is no phase boundary between weakly and strongly disordered Mott insulators.\\

\section{Conclusions and outlook}

In this paper we discussed the properties of correlated lattice fermions in the presence of disorder acting only on spin-up particles. We computed the paramagnetic ground state phase diagram and one-particle quantites such as the spin resolved spectral functions and the particle densities per spin in the extended states. Furthermore, we identified three different phase transitions which separate four distinct phases: a correlated metal, a disordered Mott insulator, a spin-selective localized phase, and a coexistence regime. In the  spin-selective localized phase  at strong disorder and weak interaction the  spin-up particles are localized, whereas spin-down particles are itinerant.

The phase diagram and the properties of such a correlated disordered system can be studied experimentally when a polarization dependent disorder is applied to an optical lattice filled with fermionic cold atoms.

In the future it will be interesting to investigate also antiferromagnetic long range order, the effect of temperature, and the effect of a gradual turning-on of the spin asymmetry of the disorder. Work along these lines is in progress.

This work was supported by the Foundation for Polish Science (FNP) through the TEAM/2010-6/2 project,
co-financed by the EU European Regional Development Fund (JS, KB). Support by the Deutsche Forschungsgemeinschaft through TRR 80 (DV, KB) is also acknowledged.

%\newpage

\end{document}